# Local spin dynamic arising from the non-perturbative SU(2) gauge field of the spin orbit effect

S. G. Tan,[1] M. B. A. Jalil,[2] Xiong-Jun Liu,[3]

[1]Data Storage Institute, DSI Building, 5 Engineering Drive 1, (Off Kent Ridge Crescent, National University of Singapore) Singapore 117608

[2]Information Storage Materials Laboratory, Electrical and Computer Engineering Department, National University of Singapore, 4 Engineering Drive 3, Singapore 117576

[3]Department of Physics, National University of Singapore, 2 Science Drive 3, Singapore 117542.

**Abstracts**

We use the non-perturbative gauge field approach to study the effects of spin orbit coupling on the dynamic of magnetic moment. We present a general equation of motion (EOM) which unifies i) the spin orbit coupling effect derived from the SU(2) spin gauge field, and ii) the moment chirality effect previously derived from the topological U(1)xU(1) rotation gauge under the adiabatic condition. We present a modified Landau-Liftshitz-Gilbert equation and discuss the implication of the modified EOM in various technological applications, such as current-induced switching and trajectory of magnetic moments in spin-valve multilayers, magnetic memory and diluted magnetic semiconductor.

(98 words)

Following the technological development of giant magneto-resistance (MR) which utilizes external magnetic fields for moment switching, it has been of interest lately to investigate the dynamic of magnetic moment interacting with spin-polarized current. The concept of spin torque, or spin transfer was first advanced [1,2] to explain a phenomenon known as current induced magnetization switching (CIMS) [3-6] in magnetic memory. Various transport theory [7-13] has of late been deployed to study CIMS in magnetic multilayers. Recently, use has been made of the continuity of spin flux (a non-conserved quantity) to derive the self-consistent equation of motion (EOM) for magnetic moment [14]. The microscopic approach [15-18] has also yielded higher order terms in the EOM as well as given new insights to magnetic moment dynamics and its technological implications. The dynamic of magnetic moment is also important in the spin valve recording device as it has direct implication to electrical signal disturbance. All of these have led to renewed interest in the EOM beyond the conventional Landau-Liftshitz-Gilbert (LLG) description.

In this article, we study magnetic moment dynamic under the influence of spin orbit coupling, which is present in semiconductor with bulk/structural inversion asymmetry as well as rare-earth magnetic materials. We show that in multilayer magnetic structure, stray electric fields as well as externally applied electric fields in specific directions could also have an effect on the EOM under the condition that electron spin aligns adiabatically with the local moment. In our work, we use the non-perturbative quantum field approach that describes electron motion in terms of gauge fields. This is because spin orbit coupling and non-trivial magnetic moment chirality result in electron motion [15-19, 21] that follows the typical trajectory of a charged particle under the influence of spin-dependent Lorentz force. The resulting moment dynamic is proportional to the energy gradient with respect to magnetization, which is given by the inner product of the gauge field and the current density. Since the gauge field derives its form from the Dirac spin orbit coupling term, the energy change would thus be related to the spin orbit coupling's effect on the EOM of magnetic moment.

The Hamiltonian of a system with spatially distributed magnetic moments and spin orbit coupling can be written as follows:

$$H = \sum_{k=x,y,z} \frac{1}{2m}\left(p_k + e\left[\frac{\hbar}{4mc^2}U\sigma_i E_j \varepsilon_{ijk}U^+ + \frac{\hbar}{e}U\partial_k U^+\right]\right)^2 + \frac{eg\hbar}{4m}\sigma_z M_z \quad (1)$$

where $U$ is the rotation matrix in spin space. The last term containing $\sigma_z M_z$ represents the alignment of electron spin with the local moment in the adiabatic approximation [15, 21]. The net chirality of the magnetic moments gives rise to a topological term of $U\partial_k U^+ \equiv A_k^M$. In the adiabatic limit, $A_k^M \approx \sigma_z a_k$ where $a_k = \left(A_k^M\right)_{11}$ denotes a matrix element of the monopole gauge. As for the spin-orbit coupling term, after application of the $U$ matrix, we obtain the SU(2) gauge reminiscent of the Yang-Mills gauge [20, 21, 23] of the form:

$$A^{SU(2)} = \frac{\hbar \sigma_z^r}{4mc^2}\left[\left(n_y E_z - n_z E_y\right)\hat{i} + \left(n_z E_x - n_x E_z\right)\hat{j} + \left(n_x E_y - n_y E_x\right)\hat{k}\right] \quad (2)$$

where $(n_x, n_y, n_z) = (\sin\theta\cos\phi, \sin\theta\sin\phi, \cos\theta)$. In the rotated frame, $\sigma_z \to U\sigma_z U^+ = \sigma_z^r n_z + p$, $\sigma_x \to U\sigma_x U^+ = \sigma_z^r n_x + q$, $\sigma_y \to U\sigma_y U^+ = \sigma_z^r n_y + r$, where $p$, $q$, $r$ are some 2x2 matrices, which are linear combinations of $\sigma_x$ and $\sigma_y$ matrices. We take $\sigma_z^r$ as the new reference axis, i.e. $\sigma_z^r = \begin{pmatrix} 1 & 0 \\ 0 & -1 \end{pmatrix}$. In the adiabatic approximation and reasonably weak spin orbit effect, the spin is primarily aligned along $\sigma_z^r$, and the electron exists in a state $|\varphi\rangle_r$ which is the eigensolution of $\sigma_z^r$. The expectation value of the SU(2) gauge is then given by $_r\langle\varphi|A^{SU(2)}|\varphi\rangle_r$, where the contributions from matrices $p$, $q$, and $r$ become zero.

The additional gauge potentials $A^{SU(2)}$ and $A^M$ in the presence of spin-orbit coupling and chiral magnetization, respectively, give rise to an electromagnetic interaction between the current and the magnetic moments, and result in an additional energy density term [15]:

$$\begin{aligned} E_{int} &= \langle j_\mu A_\mu \rangle \\ &= G\left[ j_x\left(n_y E_z - n_z E_y + ga_x\right) + j_y\left(n_z E_x - n_x E_z + ga_y\right) + j_z\left(n_x E_y - n_y E_x + ga_z\right) \right] \end{aligned} \quad (3)$$

where $G = \hbar/4mc^2$, $g = 4mc^2$, and $A_\mu = A_\mu^{SU(2)} + A_\mu^M$. The magnetic moments will move in order to reach the minimum energy state. As in conventional micromagnetic analysis [11, 22], the moments can be regarded as aligning themselves along an effective magnetic field, which is given by the energy gradient with respect to moment, i.e.:

$$\begin{aligned} H &= \frac{1}{\mu_0 M} \frac{\partial E_{int}}{\partial \tilde{n}} \\ &= \frac{G}{\mu_0 M} \frac{\partial}{\partial \tilde{n}} \left( j_x\left[n_\mu E_\nu \varepsilon_{x\mu\nu} + ga_x\right] + j_y\left[n_\mu E_\nu \varepsilon_{y\mu\nu} + ga_y\right] + j_z\left[n_\mu E_\nu \varepsilon_{z\mu\nu} + ga_z\right] \right) \end{aligned} \quad (4)$$

where $M$ is the magnetic moment density, $\tilde{n}$ is its direction, and $\mu_0 = 4\pi \times 10^{-7}$ TmA$^{-1}$. If we consider the low-damping limit, the magnetic moments will precess about the effective field, i.e. the general EOM of the magnetic moments can be written as $\partial_t \tilde{M} = \gamma(\tilde{M} \times \tilde{H})$ where $\gamma$ is the gyromagnetic ratio (in units of mA$^{-1}$s$^{-1}$). The precessional motion is reminiscent of $\partial_t \tilde{S} \propto \tilde{S} \times \tilde{H}$ which arises due to the non-commutative spin algebra of $\langle [S, H] \rangle$. From Eq. (4), we thus obtain the EOM as:

$$\partial_t M = \frac{\gamma G}{\mu_0} \tilde{n} \times \frac{\partial}{\partial \tilde{n}} \left( j_\lambda \left[ n_\mu E_\nu \varepsilon_{\lambda\mu\nu} + ga_\lambda \right] \right). \quad (5)$$

We will now focus on the vector calculus method to evaluate the contribution due to the SU(2) gauge component. The contribution from the U(1) gauge component (second term of the rhs of Eq. 5) has been derived in previous works and the final result is presented without proof in Eq. (6) following Ref. [2]:

$$\partial_t M = \frac{\gamma G}{\mu_0} \tilde{n} \times \left[ j_\lambda \frac{\partial}{\partial \tilde{n}} n_\mu E_\nu \varepsilon_{\lambda\mu\nu} + j_\lambda g \left( \tilde{n} \times \partial_\lambda \tilde{n} \right) \right]. \tag{6}$$

Rewriting Eq. (6) in compact tensor form, we obtain:

$$\partial_t M = \frac{\gamma G}{\mu_0} j_\lambda n_\alpha \left[ \frac{\partial n_\mu}{\partial n_\beta} \delta_{\mu\beta} E_\nu \varepsilon_{\lambda\mu\nu} + g \left( n_r \partial_\lambda n_s \hat{e}_\beta \varepsilon_{rs\beta} \right) \right] \hat{e}_\gamma \varepsilon_{\alpha\beta\gamma}, \tag{7}$$

where $\delta_{\mu\beta}$ is the Kronecker delta. Equation (7) which shows the general EOM of magnetic moment in the presence of weak spin orbit coupling and chiral magnetization, is the main result of this paper.

We now focus on possible implications of Eq. (7) in realistic condensed matter system. We consider three specific systems where the magnetization dynamics described by the EOM is prominent: 1) magnetic multilayer in the presence of large external electric $E$ fields, 2) dilute magnetic semiconductors (DMS) in the form of a 2DEG (with strong Rashba effect), and 3) bulk heavy magnetic materials. Since the effect of spin orbit coupling is additive to the effect of moment chirality, we will analyze the effect of spin orbit coupling only and in the specific case of one-dimensional current flow $j_x$. Figure 1(a) is a schematic diagram of a magnetic multilayer structure conventionally used in devices such as spin valve recording head of the hard disk drive and magnetic tunnel junction of magnetic random access memory (MRAM). Figure 1 (b) illustrates a DMS [24] in the form of a two-dimensional-electron gas (2DEG) structure of a III-V high-electron-mobility (HEMT) device. The structural inversion asymmetry due to band-bending at the heterojunction of the HEMT results in a large Rashba spin orbit effect.

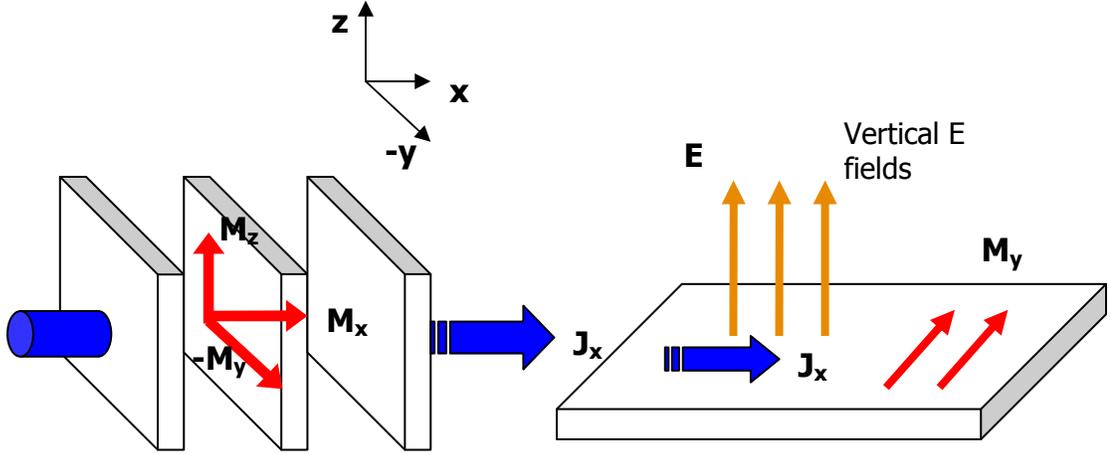

FIG. 1. (a) Magnetic multilayer with in-plane ($M_y$, $M_z$) or out-of-plane ($M_x$) magnetization, and current passing in a perpendicular-to-plane through the multilayer; (b) 2DEG of a HEMT made of DMS material with in-plane magnetization ($M_y$) and current passing in-plane along the x direction. The *E* field due to structural inversion asymmetry aligns along the z direction, resulting in spin orbit coupling.

In the above systems, the H field due to spin orbit coupling and $j_x$ is:

$$H = \frac{Gj_x}{\mu_0 M}\left(\frac{\partial}{\partial n_x}(n_y E_z - n_z E_y)\hat{i} + \frac{\partial}{\partial n_y}(n_y E_z - n_z E_y)\hat{j} + \frac{\partial}{\partial n_z}(n_y E_z - n_z E_y)\hat{k}\right), \qquad (8)$$

which simplifies to $H_y = \frac{Gj_x E_z}{\mu_0 M}$ and $H_z = \frac{-Gj_x E_y}{\mu_0 M}$. For illustration, we perform the numerical calculations in SI units (the respective SI units of $E_{int}$, H, M, are Jm$^{-3}$, Am$^{-1}$ and Am$^{-1}$). The results are plotted in Fig. 2.

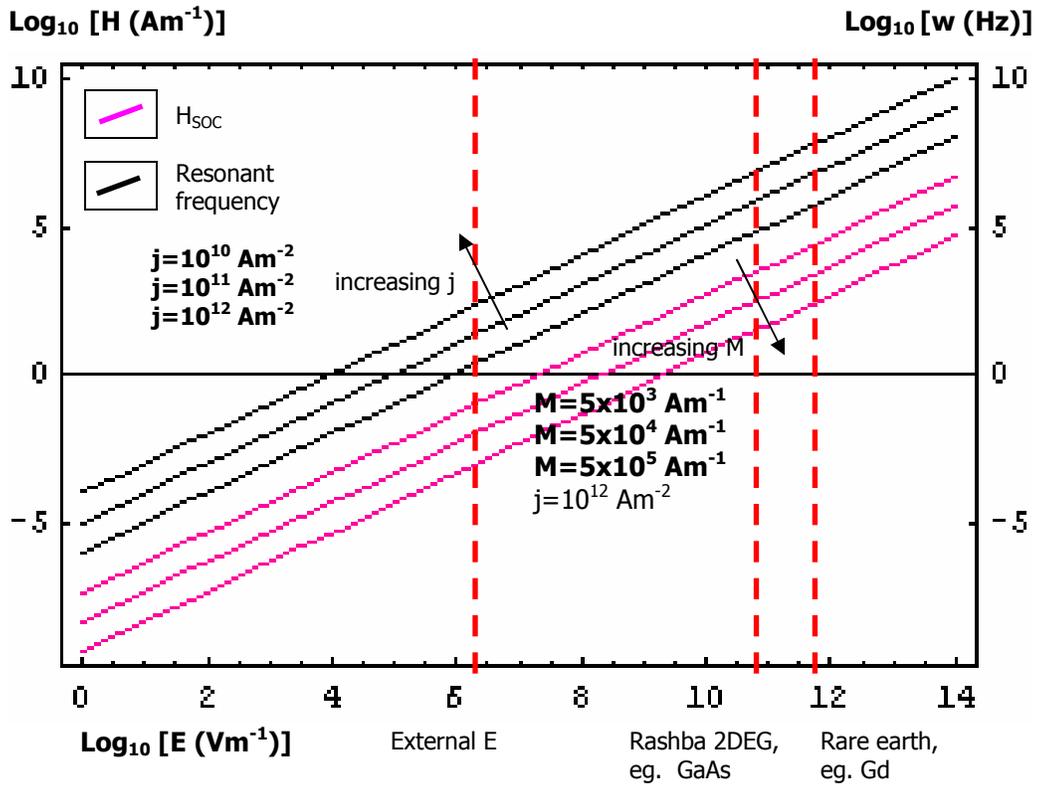

Fig. 2. Left axis: Effective field strength resulting from the SOC effect. Decreasing moment density favors high $H_{SOC}$. Right axis: Resonant frequency of the magnetic system, which increases with increasing E field strength, either applied externally or originating from internal structure.

The left axis of Fig. 2 shows the numerical values of $H_{SOC}$ for a moderately high value of $j_x=10^{12}$m$^{-2}$ for the range of E fields considered. For E field strength corresponding to the 2DEG Rashba regime, $H_{SOC}$ could reach a reasonably high value of $10^5$ Am$^{-1}$. (For comparison, the switching fields for common ferromagnetic metals like Fe, Ni and Co are respectively: 565 Oe (45x10$^3$ A/m ), 233 Oe (18.5x10$^3$ A/m), and 7429 Oe (591x10$^3$ A/m)). Additionally, in some rare earth ferromagnetic metals e.g. Gd, where the internal E field could reach $10^{13}$ Vm$^{-1}$, the corresponding effective field from the SOC effect is large. Thus, magnetic multilayers made from these rare earth elements might constitute a suitable candidate for magnetic devices which utilizes $H_{SOC}$ to achieve CIMS at reduced threshold current density.

It is worth noting that the effect of spin transfer switching due to SOC elucidated here is in addition to the other previously described sources of spin transfer torque such as the *s-d* coupling between magnetic moments and spins of conduction electrons. In the low E field regime where $H_{SOC}$ is correspondingly low (less than 1 Am$^{-1}$), there is insufficient field strength to effect moment switching. The calculated results of Fig. 2 also show that low moment density generates higher $H_{SOC}$. This is because the interaction energy is a function of the moment direction but not its magnitude. This means that in DMS materials where the saturation magnetization M is low, the effect of $H_{SOC}$ on moment switching should be even more significant.

We now investigate the trajectory of the precessional motion arising from $H_{SOC}$ and the corresponding resonant frequency of the magnetic system. It can be derived from Eq. (7) that the EOM is:

$$\partial_t \tilde{M} = \frac{\gamma G}{|M|\mu_0} j_x \left[ \left( E_z \hat{k} + E_y \hat{j} \right) M_x - \left( E_y \hat{i} \right) M_y - \left( E_z \hat{i} \right) M_z \right], \tag{9}$$

which shows that internal or externally applied electric field aligned along the in-plane magnetic moment will have an effect on its dynamics. Note that for the out-of-plane moment ($M_x$), its dynamics is only affected by the electric field orthogonal to it. From Eq. (9), we find that

$$\frac{\partial^2 M_x}{\partial t^2} = -\left( \frac{\gamma G j_x}{|M|\mu_0} \right)^2 M_x \left( E_z^2 + E_y^2 \right),$$

which yields the solution

$$M_x(t) = \left[ \frac{i\omega M_x^0 + \dot{M}_x^0}{2i\omega} e^{i\omega t} + \frac{i\omega M_x^0 - \dot{M}_x^0}{2i\omega} e^{-i\omega t} \right],$$

with the resonant frequency of the magnetic system given by:

$$\omega = \frac{\gamma G}{|M|\mu_0} j_x \sqrt{E_z^2 + E_y^2}, \tag{10}$$

where $\gamma/\mu_0 = 1.76 \times 10^{11} s^{-1} T^{-1}$, and $M_x^0$ and $\dot{M}_x^0$ are, respectively the initial $M_x$ and $\dot{M}_x$ values. By solving for the $M_y$ and $M_z$ components, the full trajectory of M is found to be:

$$\tilde{M}(t) = \left( \hat{i} + \frac{E_y e^{i3\pi/2}}{\sqrt{E_z^2 + E_y^2}} \hat{j} + \frac{E_z e^{i3\pi/2}}{\sqrt{E_z^2 + E_y^2}} \hat{k} \right) M_x(t). \qquad (11)$$

Note that when current flow is in the $j_x$ direction, the oscillation in M is strongest in the out of plane direction $(M_x)$. Even in the case where the magnetization is purely in the in-plane direction $M_y$ or $M_z$, non-vanishing out-of-plane $M_x$ component may be induced by thermal effects, and initiate the oscillatory behavior. It is important to note here that under $j_x$, the initial values of $M_x^0$ and $\dot{M}_x^0$ are crucial to the start of M oscilaltion. Equations (10) and (11) show that both the trajectory and resonant frequency can be modulated by means of the electric field strength as well as the current density. By considering the right axis of Fig. 2, we obtain the variation of the resonant frequency with $j_x$ and $E_z$ field strength. The $E_z$ field may be externally applied or arise from the internal bandstructure, as in the 2DEG structures. Our calculations show that for externally applied E-field the resonant frequency is in the kHz range, while for the larger intrinsic E-fields within the 2DEG, the resonant frequency can be as high as the microwave (GHz) range, depending on the current density. The resonant frequency can be readily tuned by changing the external or internal E-field (e.g. by a gate bias) or by modifying the current density. It can also be implied from Eqs. (10) and (11) that in the presence of indeterministic (random) time-dependent E fields, the additional spin orbit coupling term in the EOM may give rise to a new source of moment fluctuations, and thus an additional contribution to the total noise [25] in spintronic devices such as the spin valves.

In conclusion, we have modified the LLG equation by formally incorporating the effect of spin orbit coupling. We obtained a general EOM of the magnetic moments by deriving the SU(2) spin gauge field due to spin orbit coupling, in addition to the topological U(1)xU(1) rotation gauge derived previously in the presence of moment chirality under the adiabatic condition. We have also discussed the implications of our theory in various applications, e. g. current induced switching, tunable magnetization trajectory and resonant frequency, and noise. For illustration, numerical analysis has been carried out for a range of E fields that correspond to various device structures where spin-orbital effects can be significant e.g. transition and rare earth materials in magnetic multilayers, as well as DMS material in 2DEG structures.

**Acknowledgement**

We would like to thank the Agency for Science and Technology (A-STAR), Singapore for supporting theoretical work on spintronic and magnetic physics. We thank H. K. Lee, for useful discussion on the magnetization trajectory.